\documentclass[twocolumn,preprintnumbers,amsmath,amssymb]{revtex4}
\usepackage[version=3]{mhchem}
\usepackage{balance}
\usepackage{graphicx} 
\usepackage{lastpage}
\usepackage{colordvi}
\usepackage[format=plain,justification=raggedright,singlelinecheck=false,font=small,labelfont=bf,labelsep=space]{caption}
\usepackage{float}
\usepackage{here}
\usepackage{bm}

\begin{document}

\title{Clogging and avalanches in quasi-2D emulsion hopper flow}
\author{Xia Hong$^1$}
\email{hxamy728@gmail.com}
\altaffiliation[Present address: ]{LinkedIn, Sunnyvale, CA 94085, USA}
\author{Kenneth W.~Desmond$^1$}
\altaffiliation[Present address: ]{ExxonMobil, Annandale, NJ 08801, USA}
\author{Dandan Chen$^{2,3}$}
\author{Eric R.~Weeks$^1$}
\email{erweeks@emory.edu}
\affiliation{$^1$~Department of Physics, Emory University, Atlanta, GA 30322, USA}
\affiliation{$^2$~State Key Laboratory of Radiation
Medicine and Protection, School of 
Radiation Medicine and Protection, Soochow University, Suzhou,
China}
\affiliation{$^3$~School of Radiation Medicine and Protection,
Medical College of Soochow University, Suzhou, China}

\date{\today}

\begin{abstract}
We experimentally and computationally study the flow of a
quasi-two-dimensional emulsion through a constricting hopper shape.
Our area fractions are above jamming such that the droplets
are always in contact with one another and are in many cases
highly deformed.  At the lowest flow rates, the droplets often
clog and thus exit the
hopper via intermittent avalanches.  At the highest flow rates, the
droplets exit continuously.  The transition between these two types
of behaviors is a fairly smooth function of the mean strain rate.
The avalanches are characterized by a power law distribution of
the time interval between droplets exiting the hopper, with long
intervals between the avalanches.  Our computational studies
reproduce the experimental observations by adding a flexible
compliance to the system (in other words, a finite stiffness of the
sample chamber). The compliance results in continuous
flow at high flow rates, and allows the system to clog at low flow
rates leading to avalanches.  The computational results suggest
that the interplay of the flow rate and compliance controls the
presence or absence of the avalanches.
\end{abstract}


\maketitle

\section{Introduction}
\label{Introduction}

Many slowly strained materials exhibit
intermittent flow behavior: long still periods
punctuated by rapid avalanches where material flows
\cite{rbenzi2014,aGopal1995,mDennin1997,sTewari1999,jSethna2001}.
Examples include diverse phenomena such as earthquakes
\cite{dFisher1997,kDahmen1998}, general deformations of solids
\cite{kDahmen2009}, stick-slip friction due to granular layers
\cite{jGollub1997,nasuno98,jKrim2011}, 
Barkhausen noise in magnetic materials \cite{urbach95},
and sheep herded through
constrictions \cite{iZuriguel2014}.  For athermal soft materials,
avalanches are
seen in slow flows of materials such as emulsions
\cite{rbenzi2014}, bubble rafts \cite{mDennin2004}, foams
\cite{dDurian1997,sTewari1999,ePratt2003,yBertho2006,mLundberg2008},
and granular materials
\cite{rHartley2003,aJanda2008, janda09,aAguirre2010, nhayman2011,
pLafond2013,tWilson2014}.  These soft materials typically have 
amorphous structure, necessitating that flow and rearrangements are
disordered on a microscopic scale.  The slow flow speed is 
a key feature: for example, a rotating drum experiment with sand
inside demonstrated avalanches at low rotation rates and smooth
flow at high rotation rates \cite{jRajchenbach1990}.  For granular
materials, static friction can prevent the material from flowing
and can lead to avalanches. 
In systems composed of fluids such as
foams and emulsions, stresses are supported not by static friction
but rather surface tension, which resists the deformation of the
bubbles or droplets.

\begin{figure}
\centering
\includegraphics[width=8cm,keepaspectratio]{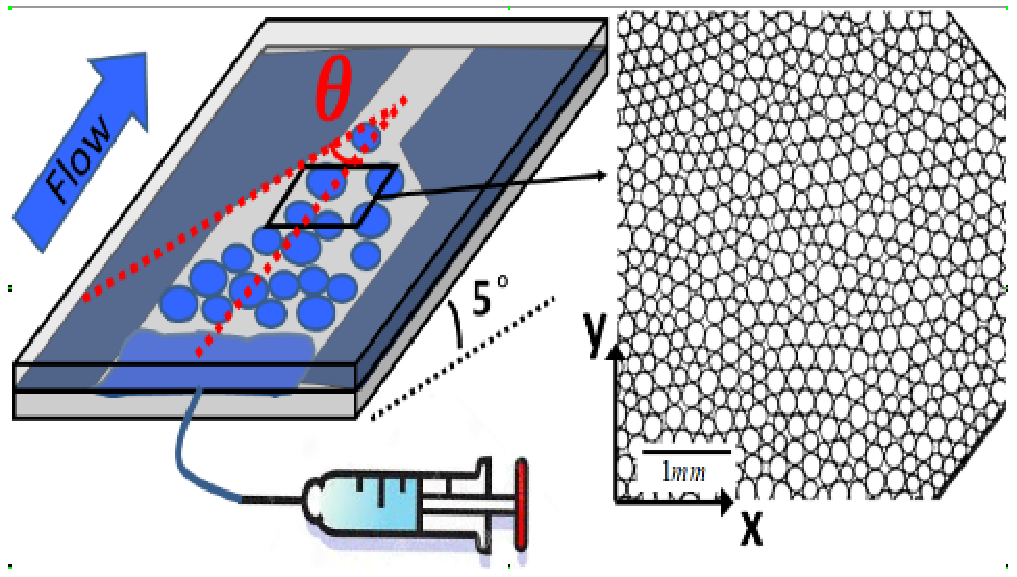} 
\caption{Schema of our sample chamber (left)
and raw image of the emulsion flowing in the $+x$-direction (right).
The hopper angle $\theta= 54^\circ \pm 5^\circ$.}
\label{chamber} 
\end{figure}

Hopper flow is a useful case study for these types of flowing
particulate materials.  In this geometry (Fig.~\ref{chamber}),
the material starts in a wide channel but then exits the chamber
through a narrow orifice.  This is of industrial interest for
storage of granular materials \cite{jenike67,nedderman82} and has
been long studied scientifically.  For example, an early paper
in 1929 examined hopper flow of various granular materials and
observed that flow halted when the exit orifice diameter was less
than about 4 particle diameters \cite{deming29}, which has been
observed many times since \cite{beverloo61,brown58,to01,tang09}.
Subsequent work found that for small exit orifices, the flow
rate fluctuates as small arches form and break near
the exit \cite{fowler59,hong92}.  For larger exit orifices,
the flow rate is smooth and generally a simple
function of the orifice size and various material parameters
\cite{brown58,beverloo61,nedderman82}.

In this manuscript, we present experimental and computational
studies of hopper flow of emulsion samples.  Our experimental
emulsions are oil droplets in water and are compressed between
two parallel glass plates so that the droplets are deformed into
pancake-like disks.  The area fractions are all above jamming
\cite{dDurian1997}.
Our exit orifices
are all small ($\sim 4$ droplet diameters across).  We drive the flow
with a pump, and given that our droplets are deformable, they cannot
permanently clog at the exit.  We see a range of flow behaviors.
At the slowest flow rates, the flow pauses for long periods of time
broken up by large avalanches of rearrangements.  At higher flow
rates droplets exit
continuously.  Intriguingly, the transition between the two flow
behaviors occurs fairly smoothly as the flow rate is increased, and
at moderate flow rates we see an intermediate type of flow behavior.
The non-constant flow seen in the experiment is in contrast with
the constant flux driving condition at the pump, indicating that
the system has compliance:  rather than being infinitely rigid,
the system expands under pressure.  Our computational studies address
this using the ``Durian bubble model'' \cite{durian95,sTewari1999},
modified to mimic our experiment and with the effects of an added
compliance.  The simulations show the same results as the experiment:
at lowest driving, the simulated compliance results in clogging
and avalanches; for larger driving, we see a smooth transition to
continuous flux.  Our results highlight the interesting influence
of compliance on the behavior of these flowing soft particles.

\section{Methods}
\label{sectionMethod}

\subsection{Experimental samples and sample chambers}
\label{subSectionDesign}

Our emulsions are mineral oil droplets in water using Fairy
detergent (mass fraction 0.025) as a surfactant to prevent
coalescence of the droplets \cite{kdesmond2013,dchen2012}.
The droplets are produced using a standard co-flow micro-fluidic
technique \cite{martinez2008}.  
The radius polydispersity of our
droplets is 1\% (standard deviation divided by mean).
To prevent droplets from organizing into crystalline arrays, for
each experiment we make a bidisperse emulsion by mixing together
two separate batches of monodisperse droplets at a volume ratio of
about 1:1.  While each individual batch of monodisperse droplets has
a low polydispersity, there is some variability between batches.
The mean diameter of the large droplets is $270 \pm 50~{\mu}$m and
of the small droplets is $200 \pm 40 ~{\mu}$m, and the diameter ratios
of the bidisperse mixtures we form are in the range $d_L/d_S =
1.5 \pm 0.2$.

In our experiment, we confine droplets between two 25 mm
$~{\times}$ 75 mm glass slides. The slides are separated by
pieces of $100~{\mu}$m transparency film sealed with epoxy.
These pieces of film act as spacers and thus create a gap
between the slides.  This gap ranges from 115 to 140 $~{\mu}$m in
different experiments. This range is mainly due to the different
amount of epoxy applied when making each chamber.  Nonetheless,
within a given sample chamber, this gap is constant with uncertainty
$1.8\%$ within any given sample chamber so the slides are parallel
(the corresponding maximum angle between two slides is less
than $1^{\circ}$).  Sample chambers for which this was not true
were discarded.  While the gap thickness varies from experiment to
experiment, our prior work found that the thickness was unimportant
as far as the contact forces droplets exert on one another when
they contact \cite{kdesmond2013}.  In all cases, the diameters
of the oil droplets are chosen to be larger than the gap of the
sample chamber. Thus, the droplets are squeezed between the two
glass slides to achieve a quasi-2D system.

The left panel in Fig.~\ref{chamber} shows the schema of the chamber.
The pieces of film are cut to form a symmetric hopper channel
with angle $\theta = 54\pm 5^{\circ}$ (see Fig.~\ref{chamber}) and
opening width $0.7-1$ mm.  The sample chamber is tilted at an angle
$5\pm 1^{\circ}$ relative to the horizontal, to use the buoyant
force of the droplets to balance the viscous friction between droplets and
glass slides at intermediate flow rates.  
The buoyant force is due to the density difference
between water and mineral oil ($\rho_{\rm water}= 1.00 $~g/cm$^3$,
$\rho_{\rm oil}= 0.83 $~g/cm$^3$).  First we load the emulsion into
the sample chamber, and then behind the emulsion we add pure
mineral oil.  A syringe pump injects additional mineral oil into
the chamber at constant flux rate to push the emulsion through the
chamber and thus funnel the droplets through the hopper exit.
The syringe pump is connected to the chamber via Teflon
tubing.

We use a microscope with a 1.6$\times$ objective lens to image the system,
focusing on the chamber midplane where the 2D droplet images are clearest.
A CCD camera records the images in the region close to (0.5-2 mm away from) the hopper
opening. Depending on the mean
speed of the flow in a given experiment, the camera frame rate
is between 0.2 and 2 images/second. This is sufficient to track
the trajectory of each individual droplet using standard
software \cite{crocker96}, even
at the maximum velocity $0.06 \langle D \rangle$/s, where
$\langle D \rangle$ is the mean diameter of the droplets. The
right panel in Fig.~\ref{chamber} shows a typical raw image,
in which we record hundreds of droplets within the field of view.
Typically, we have 100-200 droplets in the
field of view.  In the 45 experiments, an average of 425 droplets are
seen to exit during an experiment, although the
exact amount varies from $\sim 100$ to $\sim 1000$.

\subsection{Experimental control parameters}
\label{subSectionControlParameters}

One of our main control parameters is the area fraction
$\phi$
occupied by oil droplets, as measured
from our image analysis.  $\phi$ is somewhat controllable by
what we put into the sample chamber:  ahead of time, we prepare
bulk emulsion samples at different 3D volume fractions.
All our reported $\phi$ in this paper are the
measured values from the image analysis.
From the post-processed images,
we observe that $\phi$ has only minimal fluctuations during an
experiment, with a relative standard deviation 
no more than $0.5 \%$.  
These fluctuations are primarily due to the finite field
of view, with $\phi$ changing when droplets flow in and out. 
In flowing suspensions of solid particles there can be a
self-filtration effect \cite{haw04}, but we see no evidence of
this (which would be signaled by a monotonic increase of $\phi$).
Additionally, we look for water flow relative to the emulsion
droplets \cite{fFairbrother1935,fBretherton1961}
by adding tracer particles to the water for a few cases.
In every case, the water flows at the same rate as the oil
droplets.  For example, in some situations, the oil droplets
cease flowing for a period of time, and during those times the
water is also seen to cease flowing.
There is an additional possible
systematic uncertainty for $\phi$ as the apparent size of each
droplet depends 
on the illumination settings of the microscope.  We keep these
settings constant between each experiment.

The other main control parameter for our experiments is the flux
rate $F$.  We take a total of 45 data sets with
$0.83 \leq \phi \leq 0.99$ and $0.0001 \leq F \leq 0.02$~ml/hr.  
For each experiment, $F$ is set by a syringe pump and
thus is constant at the pump.  However, the observed flow velocity fluctuates.
This is likely due to some compliance in the sample chamber,
allowing sample to flow in slightly without having to flow out, and
building up pressure until it is released by droplets flowing out.
Therefore, rather than using $F$ to parameterize the
experiments, we instead use the observed flux rate $r_{\rm
expt}$, measured by the total number of droplets that exit the
sample chamber divided by the total observation time.

\subsection{Computational methods}
\label{simulation}

As noted above, while the pump provides a constant flux rate
$F$, the observed flow velocity fluctuates due to sample chamber
compliance.  We use a simulation to better study the importance
of compliance.  In particular, we use the ``Durian bubble model,''
introduced in \cite{durian95}.  We use the version as modified
in \cite{sTewari1999} to account for variable numbers of nearest
neighbor particles and as further modified in \cite{hong17}
to account for viscous friction between the moving droplets and
the glass walls that make the experimental droplets quasi-2D.
In this model, droplets are considered as disks of fixed radius
$R_i$ (for disk $i$) with a repulsive force when they overlap,
meant to approximate the influence of surface tension for real
droplets.  Droplet motion is assumed to be at low Reynolds number
(Re $\approx 10^{-2}$ at most in our experiment), so the model
sets the droplet velocity by having all repulsive droplet-droplet
forces (or droplet-wall forces) balanced with the
velocity-dependent viscous forces.  The resulting equation is
\begin{equation}
\label{bubblemodel}
\sum_{j}[ \vec{F}^{\rm contact}_{ij} +
\vec{F}^{\rm viscous}_{ij}] +
\vec{F}^{\rm wall}_i +
\vec{F}^{\rm driving}_i +
\vec{F}^{\rm plates}_i = 0.
\end{equation}
The repulsive contact force between droplets $i$ and $j$ is given by
\begin{equation}
\label{repulsive}
\vec{F}_{ij}^{\rm contact} = 
F_0 \Big[\frac{1}{|\vec{r}_i - \vec{r}_j|} -
\frac{1}{|R_i + R_j|} \Big] \vec{r}_{ij},
\end{equation}
using the droplet radii $R_i$, their positions $\vec{r}_i$, and
the vector $\vec{r}_{ij} = \vec{r}_j - \vec{r}_i$.  The neighbors
$j$ are defined as those droplets for which $|\vec{r}_{ij}| <
R_i+R_j$, that is, overlapping circles.  The repulsive wall force
is similar, pointing away from the wall and using $R_{\rm wall} =
0$.  The viscous forces between two droplets act if they are
overlapping and moving with different velocities:
$\vec{F}_{ij}^{\rm viscous} = b_{\rm droplet} (\vec{v}_j - \vec{v}_i)$, a force
which attempts to equalize their velocities.  The viscous force
from the confining plates is given by $\vec{F}_{ij}^{\rm plates}
= -b_{\rm plate} R^2_i \vec{v}_i$.  As in prior work
\cite{hong17,tao21}, we take $F_0 = b_{\rm droplet} = b_{\rm plate} =
1$, and use droplets with a Gaussian distribution of droplet
sizes (mean diameter 1, standard deviation 0.1).  (The desire to
use a Gaussian distribution in the simulation, rather than a
bidisperse distribution to better match the experiment, is that
this way the simulations are consistent with our prior work
\cite{hong17,tao21}.  Also, the details of the particle size
distribution should not matter that much to the overall
phenomenology we're studying; the main goal in both simulation
and experiment is to avoid the particles organizing into
hexagonal crystals, as they would do in a monodisperse sample.)
The unit of time
in the simulation is $b_{\rm droplet} \langle R \rangle / F_0$, the
time it takes two droplets to push apart, limited by viscous drag.

The final force in the model is the driving force. To understand
this force we will digress to discuss the influence of
compliance on regular fluids in a microfluidic chamber, following
the argument in Tabeling \cite{tabeling10} (see also \cite{martin75,stone04}).  
Tabeling considers the
case of an incompressible fluid driven by a constant flux pump
at one end of a long tube with elastic wall compliance; the fluid
exits the tube at the other end.  Initially before the pump is
started, the system has pressure $P_0=0$ and tube diameter $D_0$.
When the pump drives the fluid, the pressure $P(t)$ at the pump
end of the tube
increases, putting stress on the tube walls.  
The hoop stress and axial stress are related to the pressure as
\begin{equation}
\sigma_{\theta\theta} = 2 \sigma_{zz} =  \frac{D}{2 T} P
\end{equation}
in terms of the tube diameter $D$ and tube wall thickness $T \ll D$
\cite{timoshenko56}.
(The radial stress is negligible in tubes \cite{timoshenko56}.)
Positive hoop stress tries to increase the tube diameter $D$, and positive
axial stress tries to increase the tube length $L$, with the
amount of increase limited by the 
Young's modulus $E$ for the tube material.  These expansion
effects are coupled via the Poisson ratio $\nu$ of the tube
material, so the changes in the tube dimensions for small
pressure increases are \cite{timoshenko56}:
\begin{equation}
\frac{\Delta L}{L} = \frac{1}{E} (\sigma_{zz} - \nu \sigma_{\theta \theta})
= \frac{D}{4 T E} (1 - 2\nu) P
\end{equation}
and
\begin{equation}
\frac{\Delta D}{D} = \frac{1}{E} (\sigma_{\theta\theta} - \nu \sigma_{zz})
= \frac{D}{4 T E} (2 - \nu) P.
\end{equation}
The fractional change in volume is given by
\begin{equation}
\frac{\Delta V}{V} \approx \frac{\Delta L}{L} + 2\frac{\Delta D}{D} =
\frac{D}{4 T E} (5 - 4 \nu) P \equiv P / S
\end{equation}
where $S \sim E$ relates to the stiffness, that is, the
resistance of the tube material to stress.
We can integrate both sides to relate the pressure $P(t)$ to the
volume $V(t)$ as
\begin{equation}
\label{tabel}
P(t) = S \ln \left( \frac{V(t)}{V_0} \right) 
\end{equation}
and we see that a change in $P$ has a larger influence on $V$
when $S$ is small.  This would be the case if the tube is made
of a more flexible material.  Note that while we have
derived this for a cylindrical tube, the relation Eqn.~\ref{tabel}
is quite general and applies for different geometries
with $S \sim E$ in all cases.  The exact relation
between $S$ and $E$ depends on the specific geometry.

We now consider how this relation between pressure,
volume, and elasticity applies to our simulation.
In two dimensions, the instantaneous area of the sample chamber
is $A(t)$, with $A(t) = A_0$ at $t=0$.  
The pump moves to try to create
a constant flux $r a_0$, but initially $P \approx P_0 =
0$ so the tube expands without any fluid flowing out of the end of
the tube.  Here $a_0 = \pi \langle R^2 \rangle$ is the mean area
of one droplet, so that $r$ is the number of droplets that should
exit the hopper per unit time.  The increasing $A(t)$ ties to an
increasing pressure $P(t)$ (at the pump) and this pressure gradient
then can push fluid out of the far end of the tube.  There are
steady-state values of $A$ and $P$ such that the flux out of the
far end of the tube is $r$.

However, the simulation considers not a regular fluid but rather
a collection of soft particles which are capable of clogging
\cite{hong17,tao21}.  In other words, even with $P > 0$ the system
may clog, causing $A$ to increase (as the pump continues moving)
and increasing $P$ via Eqn.~\ref{tabel} such that the
system eventually unclogs.  To quantify this, define $A_{\rm out}(t)$ as the
amount of material that has exited the system.  
Define
\begin{equation}
\label{excess}
A_{\rm excess}(t) = r  a_0 t - A_{\rm out}(t),
\end{equation}
the difference between the amount of fluid the pump has moved
into the tube
(area $r a_0 t$) and the amount of fluid that has actually left the tube.  
Thus, the area of fluid contained in the tube is given
at time $t$ by by
\begin{equation}
A(t) = A_0 + A_{\rm excess}(t)
\end{equation}
so that
\begin{equation}
\label{finaldeqn}
A(t)/A_0 = 1 + A_{\rm excess}(t)/A_0.
\end{equation}
This expression can be put into Eqn.~\ref{tabel} to relate $A_{\rm
out}(t)$ to the pressure $P(t)$ at the pump, causing a pressure
gradient acting on each particle.

In the simulation, we treat the pressure gradient as if it is a
gravitational force with strength $g(t)$, so that 
\begin{equation}
\vec{F}^{\rm driving}_i(t) = g(t) R_i^2 \hat{x}
\end{equation}
pushes the particles toward the hopper exit.  The choice of this
force being proportional to $R_i^2$ is for two reasons.  First,
this dependence matches that of $F^{\rm plates}$ such that an
isolated droplet moves with constant terminal velocity, as
expected.  Second, this lets us compare with our prior work which
explicitly considered gravitational forces \cite{hong17,tao21}; as will
be demonstrated, this is a fruitful comparison that will
illustrate the interplay between compliance, clogging, and
avalanches.

To put this all together, Eqn.~\ref{tabel} is rewritten
\begin{eqnarray}
\label{almost}
g(t) &=& S \ln \left( 1 + A_{\rm excess}(t) / A_0 \right)\\
&\approx&
\ln \left( 1 + A_{\rm excess}(t)[S/A_0] \right),
\end{eqnarray}
where the approximation is valid for $A_{\rm excess} \ll A_0$.
Given that $S$ and $A_0$ now appear in a ratio, we define this
ratio to be the effective stiffness $s$, and rewrite
Eqn.~\ref{almost} as
\begin{equation}
\label{gravity}
g(t) = \ln \left( 1 + s A_{\rm excess}(t) \right).
\end{equation}
In the simulation we will vary $s$
from $10^{-5}$ to $2 \cdot 10^{-3}$, and in practice this
equation will lead to values of $g$ in the range 
$g \sim 10^{-4} - 10^{-1}$, consistent with our prior work which
found that clogging occurs in this range \cite{hong17}.
$A_{\rm out}$
increases by $\pi R_i^2$ when droplet $i$ exits; a droplet that
is partially out of the hopper contributes its fractional area to
$A_{\rm out}$.

To run the simulation we put 500 particles into the hopper
near the exit and start with $A_{\rm excess}(t=0) = 0$, $g(t=0)
= 0$.  The hopper is set with a fixed opening width $w/d$ and
stiffness $s$.  $g$ then increases according to Eqns.~\ref{excess},
\ref{gravity} using the desired flux rate $r$.  Droplets that exit
the hopper add their area to $A_{\rm out}(t)$; the droplets are then
replaced into the hopper touching the droplets farthest away from
the hopper exit, so that the number of particles remains constant.
Indeed, at steady state, fluid exiting the compliant
tubing would be replaced by new fluid injected by the syringe
pump, keeping the total amount in the tubing constant. This choice
of a constant number of droplets is equivalent to saying
that $A_{\rm excess}(t)$ is always small compared to $A_0$.

We run the simulation using fourth-order Runge-Kutta to solve the
differential equations for the droplet velocities, typically using
a time step of 0.1 \cite{hong17}.  We would expect there is some
value of $g$ such that the time-averaged flux matches $r a_0$, but we
would also expect that $g$ will fluctuate and that $dA_{\rm out}/dt$
will fluctuate around the value $r a_0 $.  The data are examined and the
initial transient is discarded, such that for the remaining data
$dA_{\rm out}/dt$ indeed fluctuates around $r a_0$.  The simulation
is then run until 1000 droplets exit the hopper.

\section{Experimental Results}
\label{sectionResults}

\begin{figure*}
\centering
\includegraphics[width=14cm,keepaspectratio]{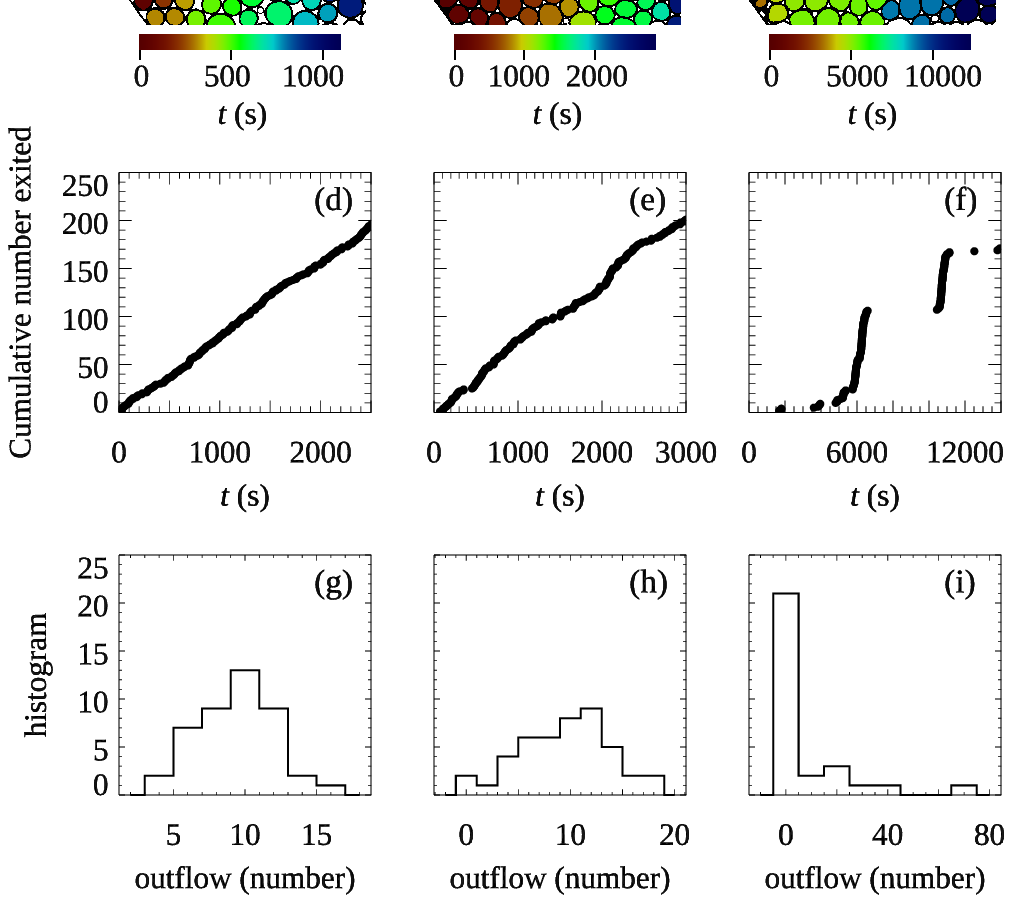}
\caption{Description of the three flow behaviors.
(a-c) Images of the samples at a particular time, with the color
indicating the time when the droplet exits (see color
bars).  The red droplets exit earlier and blue droplets exit later.
(d-f) The cumulative number of droplets that
have exited the hopper as a function of time. (g-i)
Histograms of the number of droplets exiting the hopper within a short
time window $\Delta t$, chosen such that the mean of the histogram is 10
droplets.
The flow conditions are: (a,d,g) smooth flow, $\phi = 0.87$,
$r_{\rm expt} = 0.17$~s$^{-1}$.
(b,e,h) Intermediate,
$\phi = 0.96$, $r_{\rm expt} = 0.085$~s$^{-1}$. (c,f,i)
Avalanche, $\phi = 0.96$, $r_{\rm expt} = 0.020$~s$^{-1}$.
}
\label{ava} 
\end{figure*}

We observe a wide range of flow behaviors as we vary $F$ and
$\phi$ for different experiments.  For large $F$, droplets flow
continuously and smoothly (referred as {\em smooth flow} cases).
For small $F$, we see avalanche-like flow (referred
as {\em avalanche} cases).  For intermediate flux rates $F$,
we observe {\em intermediate cases} between these two flow patterns.
As will be discussed below, we do not see any clear dependence of
these flow patterns on the area fraction $\phi$.


We summarize these three flow behaviors in Fig.~\ref{ava}.
The three pictures in Fig.~\ref{ava}(a)-(c) use color to show
the time each droplet exits the hopper opening to the right.
Red droplets exit the earliest, and blue the latest.  The left
picture is a smooth flow case, which shows a smooth gradient in
color. The right one shows an avalanche case, where droplets have
distinct groups of colors indicating that droplets exit the hopper
in bursts.  Note that the color scale of each plot corresponds to
a different amount of time, as specified by the color bar.

Figs.~\ref{ava}(d)-(f) quantify these pictures by showing the cumulative
number of droplets that have exited the hopper as a function of
time for our three flow cases.
In the smooth flow case (d), the data form a smooth curve with
a well-defined slope, showing that droplets exit the hopper
continuously at a fairly constant rate.  The intermediate case
(e) shows fluctuations in the rate, although it is still fairly
continuous. In avalanche case (f), there are stretches of time
where no droplets exit, followed by discrete sudden flow events
where many droplets exit within a short period of time, indicated
by the vertical portions of the data in (f).  Specifically, the
first vertical line at $t \approx 6000$~s relates to all of the
light green droplets in (c) that exit at nearly the same time.  Again,
the existence of avalanches despite the constant flux set by the
syringe pump shows that there is some compliance in the chamber,
such that the pressure builds up before an avalanche.

Fig.~\ref{ava}(g)-(i) show the histograms of numbers of droplets
that outflow within a short time window $\Delta t$.
$\Delta t$ is chosen to make the mean outflow size
to be 10. The smooth flow case (g) has a Gaussian shape while the
avalanche case (i) has a few rare but large events.  To quantify
this, the skewness values for these distributions are (g) 0.15,
(h) -0.03, and (i) 2.2 for smooth flow, intermediate, and avalanche
cases respectively.  Not surprisingly, the avalanche case has a
large positive skewness, and this is generally true that all
avalanche flow cases have positively skewed distributions.
Given that the avalanche cases have few events overall
($\sim 100$ in some cases), our
skewness data are noisy and we cannot resolve any clear trend in
the skewness as a function of our control parameters.  The general
picture shown in Fig.~\ref{ava}(g)-(i) is clear, though, that
avalanche cases have distributions with positive skewness and
there is a trend toward more symmetric distributions with skewness
$\approx 0$ as the mean flux rate
$r_{\rm expt}$ increases.

\begin{figure}
\centering
\includegraphics[width=6cm,keepaspectratio]{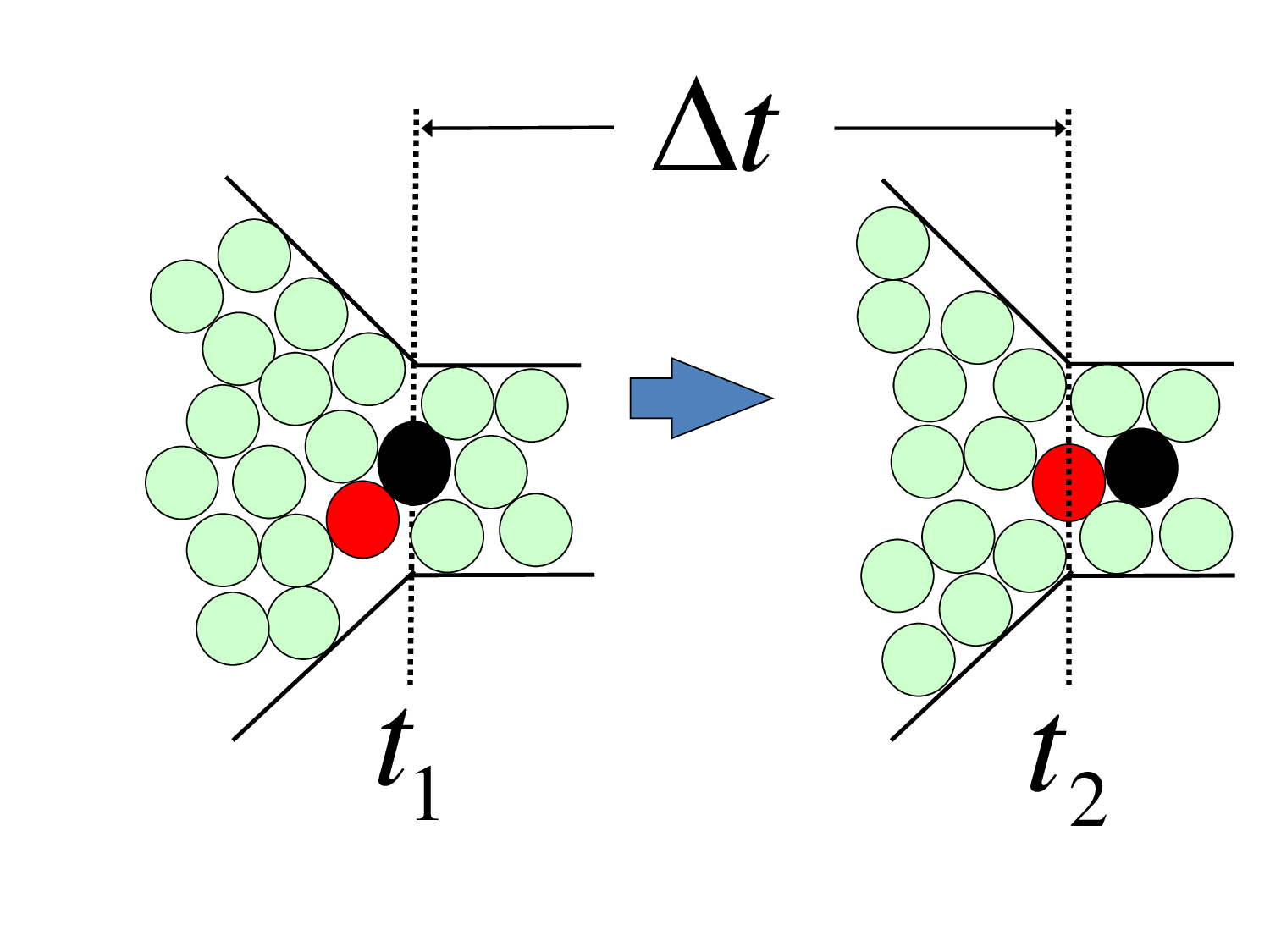}
\caption{Schema of the definition of interval
${\Delta}t$. The left figure is at time $t_1$ when the black
droplet exits the hopper. The right figure is at time $t_2$ when
the red droplet exits the hopper.}
\label{intervala} 
\end{figure}

To better quantify the difference of these flow behaviors, we
focus on the temporal behavior of the flow.
In avalanche cases, discrete sudden flow events are separated by
time intervals where droplets barely move and no droplets exit
the hopper.  Accordingly, we define the time between two successive
droplets exiting the hopper as the interval ${\Delta}t$.  As shown
in Fig.~\ref{intervala}, we set $t_{1}$ as the time when the
black droplet exits the hopper, $t_{2}$ as the time when the
next droplet (in red) exits, and then ${\Delta}t =
t_{2} - t_{1}$.  In most experiments, we observe at
least 100 droplets exit and thus have that many intervals.  For
the fastest flow rates, we have over 1000 intervals measured.

\begin{figure}[t!]
\centering
\includegraphics[width=8cm,keepaspectratio]{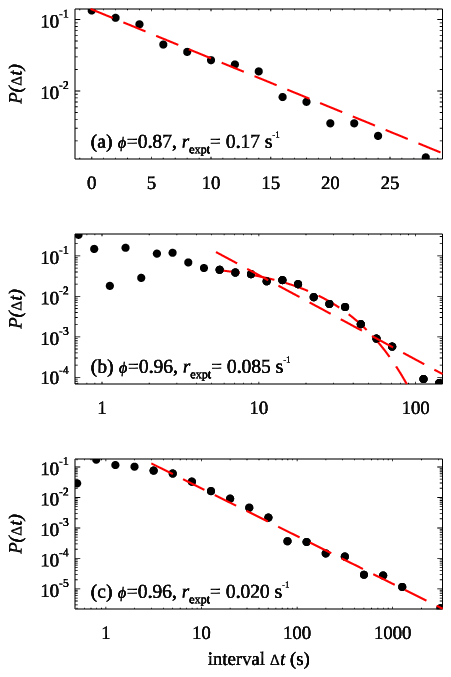} 
\caption{Typical examples of three types of probability
distribution functions of ${\Delta}t$:
(a) exponential distribution, (b) intermediate case, (c) power
law distribution.
The area fraction $\phi$ and mean flux rate $r_{\rm expt}$
are as indicated in each panel.  
In (a) the line shows an exponential fit 
$P({\Delta}t){\sim}e^{-{\Delta}t/ \tau}$ with $\tau = 6.3$~s.
In (b) the straight line is a power-law fit 
$P({\Delta}t){\sim}{\Delta}t^{-\alpha}$
with $\alpha=2.1$ and the curved line is
an exponential fit with $\tau= 12.7$~s.
In (c) the line is a power-law fit with $\alpha=1.6$.}
\label{intervalhist} 
\end{figure}

It is apparent in the plots in Fig.~\ref{ava}(d-f) that the
distributions of ${\Delta}t$ are different for the smooth flow
and avalanche cases. In smooth flow, the values of ${\Delta}t$
are small and do not fluctuate much. In the avalanche case,
${\Delta}t$ is sometimes small (vertical portions, where many
droplets exit over a short time interval) and sometimes large
(horizontal stretches, where a long time passes between one
droplet exiting and the next).  Figure~\ref{intervalhist} shows
the probability distribution functions for ${\Delta}t$ for the
same three data sets shown in Fig.~\ref{ava}. The smooth flow case
shown in Fig.~\ref{intervalhist}(a) is well fit to an exponential,
as shown by the dashed red line; note this is a semilog plot.
The exponential fit suggests
that the time between events follows a Poisson process, where
events occur continuously and independently with a constant
mean rate.  The avalanche case shown in panel (c) is well
fit to a power law, as shown by the dashed red line; note this
is a log-log plot.  The fit in this case is given by 
$P({\Delta}t){\sim}{\Delta}t^{-\alpha}$ with $\alpha=1.6$, and
the power law regime covers more than 2 decades in ${\Delta}t$ and
more than 4 decades in probability.  The tails correspond to the
long periods of time where droplets barely move.  The intermediate
case in panel (b) is plotted on log-log axes, and can be fit with
either a power law (straight line) or an exponential (curved line);
neither fit is perfect.  The exponential fit fails for the
largest ${\Delta}t$ while the power law is not adequate to
describe the small ${\Delta}t$ region.

\begin{figure}
\centering
\includegraphics[width=8cm,keepaspectratio]{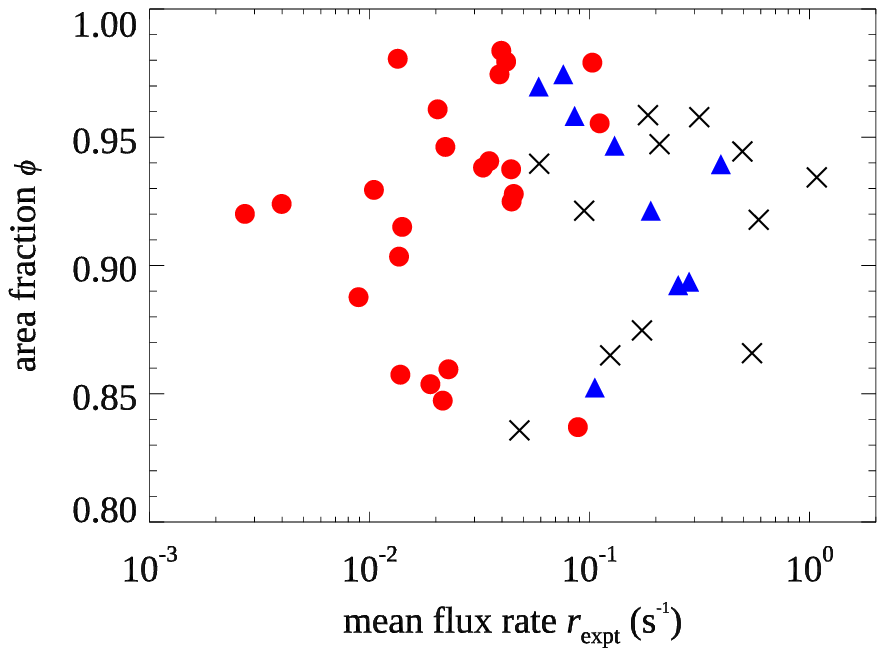} 
\caption{Phase diagram of fitting patterns of
$P(\Delta t)$
in terms of area fraction $\phi$ and flux rate $r_{\rm
expt}$ for our 45 experiments.
Red circle:  power law;
blue triangle: intermediate; black cross: exponential.}
\label{phase} 
\end{figure}

In our experiments we vary both $\phi$ and flux rate. For each experiment, we use the shape of $P(\Delta t)$
to describe its flow behavior. Figure \ref{phase} shows the
phase diagram of fitting patterns.  There is no obvious trend
with $\phi$, but more clearly a transition from avalanche flow
(red circles) to avalanche flow (black cross) with increasing
$r_{\rm expt}$.  Note that the judgment about the
best fitting function is done by eye.  The quality of each fit
depends on which range of data is used for the fit, and while
we have tried several ways to approach the fitting procedure
more systematically, none seem satisfactory for the intermediate
cases, and none affect the appearance of Fig.~\ref{phase} in any
substantial way.

\begin{figure}[tb]
\centering
\includegraphics[width=8cm,keepaspectratio]{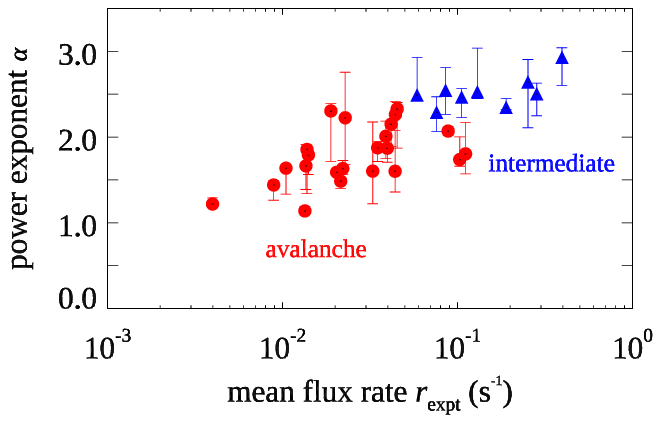} 
\caption{(a) The power law exponent $\alpha$ 
as a function of the experimental flux rate $r_{\rm expt}$.
For the power law fits, only data in the tail are used for the
fit.  Different choices of the minimum $\Delta t$ used for the
fit give rise to different values of $\alpha$, 
reflected in the error bars shown.
}
\label{intervalfitting} 
\end{figure}

The phase diagram of Fig.~\ref{phase} is perhaps unsatisfying as
the intermediate cases (blue triangles) are mixed in with the
other two cases.  However, by ignoring $\phi$ and focusing only
on the flow rate dependence, the data become more unified.  In
particular, Fig.~\ref{intervalfitting} shows the 
relation between the power law exponent of $P({\Delta}t)$ and
$r_{\rm expt}$.
The power exponent $\alpha$
increases as the flux rate increases.
Even when the power law fit is not perfect (triangles), the data
still follow the general trend started by the well-fit power law
cases (circles).  Smaller values of 
$\alpha$ indicate a broader distribution, where the large $\Delta
t$ events are more significant:  these are the avalanche cases
with long pauses between short bursts when many droplets exit.
This is similar to
previous experimental studies of sheared granular materials,
which have power law distributions of various stick-slip
event properties including forces, energy, and avalanche sizes
\cite{kdahmen2011,rCandelier2009,zhang2010,nhayman2011,tMajmudar2005,aBaldassarri2006}.
Likewise, studies of clogging with sources of vibration or
agitation find power law distributions of exit times
\cite{janda09,iZuriguel2014}.  
To comment briefly on the exponential cases, note that the
exponential fitting parameter $\tau$ corresponds to the mean
interval between droplets exiting, and thus is connected to the
flux rate by $\tau = 1/r_{\rm expt}$.

\section{Simulation results}
\label{simresults}

As described in Sec.~\ref{simulation}, the aim of the simulation
is to describe the flow of soft 2D particles with viscous
interactions and in a system with controllable compliance.  From
our prior work with fixed (gravitational) driving, we know that
such systems can clog \cite{hong17,tao21} where particles stop exiting
the hopper permanently.  This clogging occurs more easily for
narrow opening exit widths $w/d$ (in terms of the mean droplet
diameter $d$).  In the simulations we report here, the
driving can increase indefinitely
(Eqns.~\ref{excess}, \ref{gravity}) so that no clog is permanent.

\begin{figure}[tb]
\centering
\includegraphics[width=8cm,keepaspectratio]{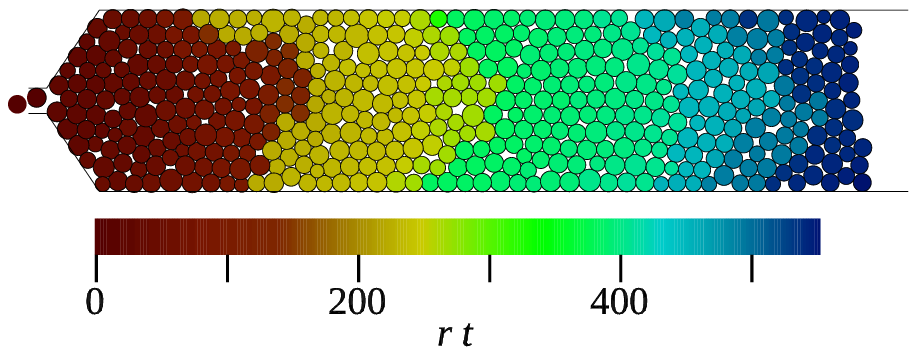} 
\caption{Image of the simulation data, colored similarly to
Fig.~\ref{ava} with color indicating the time when the droplet
exits.  For this simulation, the exit width is $w/d=1.4$ (width
divided by mean droplet diameter), the flux rate is
$r=10^{-3}$, and the stiffness is $s=3\cdot 10^{-5}$.  The exit time
distribution appears power law with exponent $\alpha=1.7$.
The total time shown is $5.53\times 10^5$, 
during which given $r=10^{-3}$ we'd expect
553 droplets to flow out; only 500 droplets flow out for
this particular time interval.
}
\label{simpic} 
\end{figure}

\begin{figure}[tb]
\centering
\includegraphics[width=8cm,keepaspectratio]{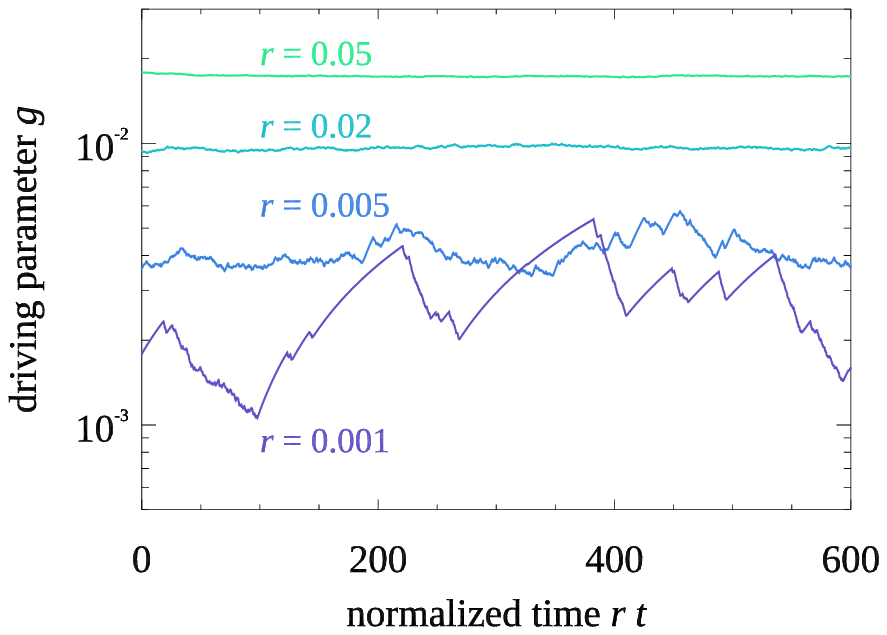} 
\caption{
The value of the simulation driving parameter $g$ as a function
of time.  Time is normalized by the rate $r$ such that on
average one droplet should exit per unit normalized time.
Stretches of data where $g$ increases are due to periods
where the system is clogged.  The
values of the rate $r$ are as indicated.
}
\label{gravtime} 
\end{figure}

Figure \ref{simpic} confirms that this simulation leads to clogging
and avalanche behavior.  In this figure the conditions have been
optimized for clogging:  a small
opening width $w/d$, small flux rate $r$, and small stiffness $s$
(see caption for details).  The behavior of $g$ is shown in
Fig.~\ref{gravtime} ($r=0.001$ data, bottom curve).  The small
flux rate and small stiffness results
in a repeated cycle where $g$ is small
enough to cause clogging, then gradually $A_{\rm excess}$ and thus
$g$ build up until the clog is disrupted and an avalanche occurs,
at which point $A_{\rm excess}$ quickly decreases, decreasing $g$
so that another clog can occur.  In Fig.~\ref{simpic} there's an
obvious division between the red and yellow droplets.  The
red droplets all exit the hopper before $rt \approx 100$.  At $rt
\approx 100$, Fig.~\ref{gravtime} shows the start of a long
nearly monotonic increase in $g$, which ends at $rt \approx 220$.
This is when the yellow droplets of Fig.~\ref{simpic} begin to
exit the hopper.

\begin{figure}[tb]
\centering
\includegraphics[width=7.5cm,keepaspectratio]{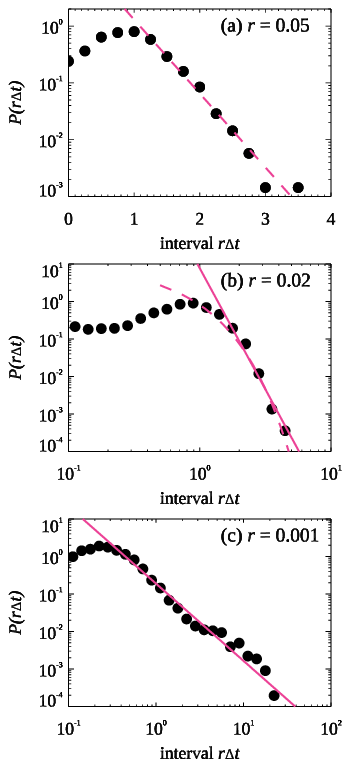} 
\caption{Typical examples of three types of probability
distribution functions of $\Delta t$ from simulation data:  (a)
exponential distribution, (b) intermediate case, (c) power law
distribution.  The flux rate $r$ is as indicated, and for these
data the stiffness is $s=10^{-4}$.  The time is normalized by
$r$, so that the mean time between droplets exiting is 1.
In (b) the straight line is
a power-law fit $P(\Delta t) \sim \Delta t^{-\alpha}$ with
$\alpha=6.5$.  In (c) the straight line is a power-law fit with
$\alpha=2.1$.  The dashed lines in (a) and (b) are exponential
fits.
}
\label{simpdf} 
\end{figure}

As with the experiment, the time $\Delta t$ between subsequent droplets
exiting the hopper varies.  Representative probability
distributions for the simulation are shown in Fig.~\ref{simpdf},
where the only parameter varied is the imposed flow rate $r$ as indicated
in each panel.  The largest flow rate corresponds to an
exponential distribution [Fig.~\ref{simpdf}(a)], the slowest
flow rate corresponds to a power law distribution
[Fig.~\ref{simpdf}(c)], and the intermediate flow rate is a bit
hard to characterize [Fig.~\ref{simpdf}(b)].

\begin{figure}[tb]
\centering
\includegraphics[width=8.5cm,keepaspectratio]{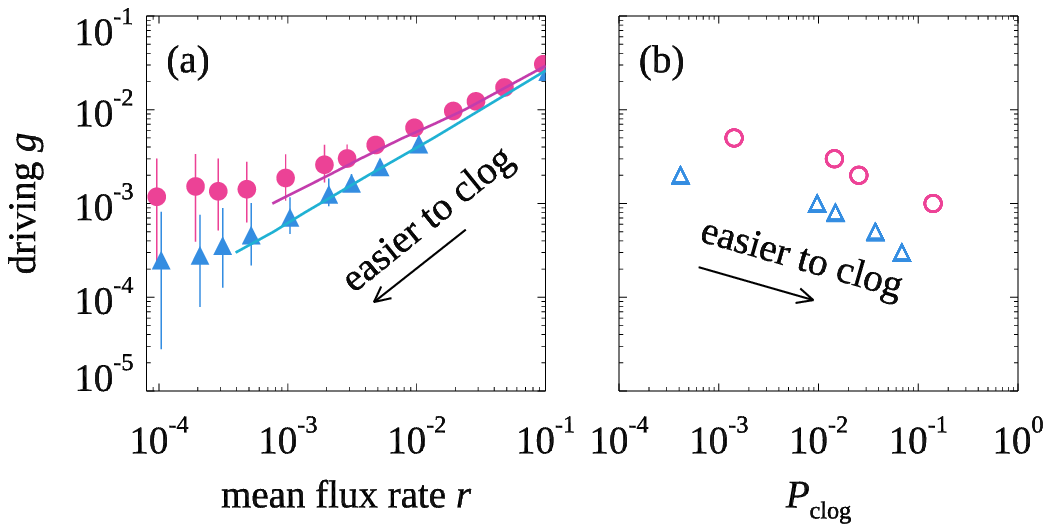} 
\caption{(a) Symbols:  the mean value of the gravitational
driving parameter $g$ as a function of the desired mean flux rate
$r$, for $w/d=1.4$ (circles) and $w/d=2.0$ (triangles).  The
thin vertical lines indicate the spread 
from the 10\%ile to 90\%ile of $g$.
Thick lines:  the mean value of the measured flux rate $r$ from
simulations with constant $g$, omitting moments when the
simulation permanently clogged.  (b) The horizontal axis shows
the probability per droplet of a clogging event from simulations
with constant gravity $g$ (vertical axis).
}
\label{gravityfig} 
\end{figure}

The simulation allows us to better understand how the avalanches
are related to clogging.  To do this we conduct a complementary set
of simulations where $g$ is fixed (in other words, an infinitely
stiff system [$s \rightarrow \infty$],
and not using Eqn.~\ref{gravity}) and the simulation is run until
clogging is observed.  As with the main simulations described
in Sec.~\ref{simulation}, we use 500 droplets and when a droplet
exits the hopper it is replaced.  When a clog occurs, the clogging
event is recorded, and the system is reset by removing the droplets
forming the clogging arch and replacing them at the back of the
hopper.  A clog is defined as when no droplets exit the hopper
for a long time, and the maximum velocity of any droplet falls
below $10^{-5}$ \cite{hong17}.  During times when clogging has
not occurred, we measure the mean flux rate $\langle r \rangle$.
These no-compliance results are shown in Fig.~\ref{gravityfig}(a)
as the solid lines -- that is, these lines are the time averaged
flux $\langle r \rangle$ emergent from the simulation at fixed $g$.
These results are comparable to the results from the first set of
simulations, plotted as symbols, where the flux rate $r$ is fixed
and we measure the time averaged $\langle g \rangle$.  
For $r > 10^{-2}$ at the right side of the graph, we get the same
results whether we fix $g$ or fix $r$.

On the left side of the graph in Fig.~\ref{gravityfig}(a), for the
simulations with compliance, the mean value of $g$ is higher than
we would expect for a given desired flux rate $r$ based on the
no-compliance simulations.  To understand this, we consider the
clogging probability measured in the no-compliance simulations.
We determine the mean number of droplets that flow
out between clogs $\langle N \rangle$, and define $P_{\rm clog} =
1/\langle N \rangle$, the mean probability of any individual droplet
clogging.  This probability is plotted in Fig.~\ref{gravityfig}(b):
to be clear, the no-compliance simulations are done with fixed
$g$ (vertical axis) and we measure $P_{\rm clog}$ (horizontal
axis), plotted this way to facilitate comparison with panel
(a).  For smaller values of $g$, clogging is easier.
Additionally, for smaller opening size $w/d$,
clogging is easier [red circles in Fig.~\ref{gravityfig}(b)].
This then explains the flux results for small gravity (small
flux) in Fig.~\ref{gravityfig}(a):  The ``expected'' value of $g$
for achieving a desired flux rate $r$ is so small that clogging
easily occurs for the given opening width $w/d$.  Thus $g$ rises
until the clog breaks, although at that point $g$ is larger than
needed for the desired flux rate $r$ and $g$ thus decreases.
It is these fluctuations in $g$ that result in long-lived clogs
and large avalanches.  Due to the logarithm in Eqn.~\ref{gravity},
the mean value of $g$ is higher than ``expected'' from simple
extrapolation of the no-compliance simulation data [the solid
lines in Fig.~\ref{gravityfig}(a)].  The thin vertical lines in
Fig.~\ref{gravityfig}(a) show the variability of $g$, highlighting
that the fluctuations are more significant when the desired
flux rate $r$ is small.  The long tails of the $P(\Delta t)$
distributions are due to unusually strong clogging arches.
This suggests that the strength of an arch -- how much
weight it can support -- may have a power law distribution as well.

Figure \ref{gravityfig}(a) also suggests that there is some
threshold arch strength that must be exceeded to break the strongest
arches.  That is, the height of the thin vertical lines is nearly
constant for low values of $r$, showing that the fluctuations in $g$
always rise above a threshold to unclog the system.  This is also
suggested in Fig.~\ref{gravtime}, where the two lowest rates $r$
have similar maximal values of $g$.  In a sense, our clogging system
is acting like a yield stress fluid \cite{coussot02}.  For a yield
stress fluid, if the applied stress is below the yield stress,
it does not flow.  The difference between a yield stress fluid
and our system is that for the former the non-flowing behavior is
a homogeneous bulk response from the entire material, whereas
in our system the clogging is due to the few specific particles
forming the arch at the exit \cite{hong17}.  For this reason,
in our simulation the particular yielding point (value of $g$
for which the system unclogs) varies from clog to clog, given
that the arch structure is variable.

\begin{figure}[tb]
\centering
\includegraphics[width=8cm,keepaspectratio]{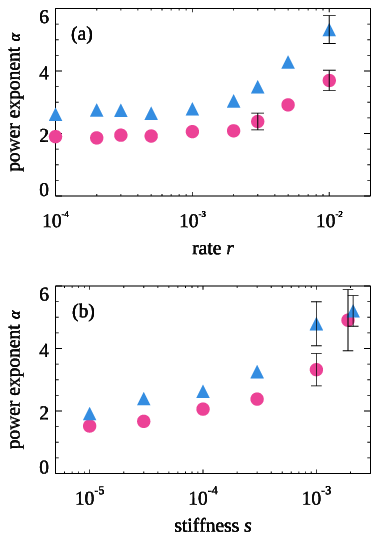} 
\caption{(a) Power-law exponent $\alpha$ as a function of flux
rate $r$ from simulation data, for $w/d=1.4$ (circles) and
$w/d=2.0$ (triangles).  The stiffness is $s=10^{-4}$.
(b) $\alpha$ as a function of stiffness $s$ for fixed flux rate
$r=10^{-3}$ and the symbols indicating $w/d$ as in
panel (a).  For both panels,
the error bars indicate the uncertainty of $\alpha$.  Where not
shown, the uncertainty is smaller than the symbol size.  
}
\label{alpha} 
\end{figure}

These results (clogging and time-varying forcing $g$) conceptually
explain the probability distribution functions for the time
intervals shown in Fig.~\ref{simpdf}.  To complete the story,
Fig.~\ref{alpha} uses the compliance simulation data to show how
the measured power law exponent $\alpha$ varies with the flux rate
$r$ at fixed stiffness $s$ in panel (a), and how $\alpha$ varies
with $s$ at fixed $r$ in panel (b).  The larger uncertainties
at $r \sim 10^{-2}$ and $s \geq 10^{-3}$ indicate that the power
law fits become dubious, signalling the crossover to exponential
distributions.  Power law exponents $\alpha > 3$ are also
reasonable indications of a transition to exponential
distributions.  The results of Fig.~\ref{alpha}
strongly suggest that indeed it is the
experimental compliance that allows for us to observe the flow rate
dependent crossover from avalanche behavior to continuous flow.
The values of $\alpha$ we see in the simulation are comparable with
the experimental values (compare Fig.~\ref{intervalfitting})
at the low end; at the high end, the simulation can generate more
data and thus measure power law exponents even for steeply decaying
functions with $3 < \alpha < 6$.

\section{Conclusions}
\label{sectionConclusions}

We have demonstrated that in a simple hopper geometry we see
behaviors changing from clear avalanches to smooth continuous flows
as we increase the mean flow rate by a factor of 100.  We quantify
these behaviors by examining the distributions of times $\Delta t$
between subsequent droplets exiting the hopper.  Intriguingly, the
transition in the flow behaviors is smooth as we increase
the flow rate:  the power law exponent characterizing the tails of
$P(\Delta t)$ smoothly varies as the flow rate increases past the
point where a power law no longer adequately describes the data
[Figs.~\ref{intervalfitting}, \ref{alpha}(a)].  One possibility
is that at any flow rate, the distribution $P(\Delta t)$ may be
describable by a power law with an exponential cutoff, and this
cutoff may smoothly move to smaller $\Delta t$ as the flow rate
increases.  However, the data we have for the intermediate cases
[such as shown in Fig.~\ref{intervalhist}(b) and
Fig.~\ref{simpdf}(b)] are hard to interpret
in the tails, and so it is difficult to resolve this question.
The rate dependence of our observations is consistent with prior
studies of athermally sheared 2D amorphous solids which demonstrated
rate dependence \cite{lemaitre09,rWhite2003,gKatgert2008,
kKrishan2008, rHartley2003, dHowell1999, cVeje1999}. A 
simulation \cite{eWoldhuis2015} based on Durian's 2D bubble
model \cite{dDurian1997} predicted a similar trend for the flow
behavior as strain rate increases.  However, this study also found a
dependence of the transition on area fraction, which we do not see.
It is likely this is due to different flow geometries (a simple
shear flow in the simulation, as compared to our hopper flow
which allows for clogging).
The dependence on velocity is also displayed in experimental
studies of sheared granular materials, where friction plays a key
role \cite{jGollub1997, nasuno98, jKrim2011}.  One of these
studies in particular also noted that more compliant driving
resulted in larger fluctuations in the sample motion
\cite{nasuno98}, in agreement with our observations. For hopper
flow in granular experiments, the presence of static friction can
make jamming and clogging obvious, where stress-supporting solid
arches form across the exit \cite{to01,hong17}.  In addition to
static friction, such experiments are also driven by a constant
force (gravity), whereas in our experiments the syringe pump
increases the pressure until flow occurs, and so no arches can
persist indefinitely.

It is also clear from the simulations that the compliance plays
an important role.  By increasing the stiffness of the system
[Fig.~\ref{alpha}(b)] we can drive the system from power-law
behavior to exponential behavior.  While our results suggest that
even for a quite stiff system there is still some hypothetical
quite slow flux rate that would lead to avalanches, it is
likely that such a low flux rate would be experimentally
challenging to control.  While our simulation method
uses some approximations, the conceptual picture is simple.  At
slow flow rates, when the sample clogs, the pressure rises until
it unclogs, and then the pressure drops to keep the mean flow
rate slow.  For fast flow rates, the driving pressure is such
that the sample never clogs, so the flow rate is steadier.

In summary, we see that the flow of an emulsion through a hopper
can vary from avalanche-like to continuous.  The transition between
these behaviors is not abrupt, but rather a continuous function of
the flow rate.  At the lowest flow rates, the power law exponent
we observe approaches $\alpha = 1$, showing that the flow has
extremely long quiescent intervals in between the avalanches.
The decrease of the power law exponent with decreasing flow rate
[Figs.~\ref{intervalfitting}, \ref{alpha}(a)] 
suggests that even with these slow
flows, we are not in a quasi-static limit, in agreement with a prior
study of slowly sheared bubble rafts \cite{Twardos05}.  In this
simple limit where the strain rate approaches zero, the flow is not
simple, but rather dominated by the rare intermittent avalanches.

\section*{Acknowledgments}

This material is based upon work supported by the National
Science Foundation under Grant No. CBET-1336401 (X.H. and K.W.D.)
and CBET-2002815 (E.R.W.)  The work was also partially supported
at an early stage by the Petroleum Research Fund (47970-AC9),
administered by the American Chemical Society.  We thank J.~Burton,
K.~Dahmen, C.~Orellana, D.~Young, and I.~Zuriguel for helpful discussions.

\end{document}